\documentclass[prb,aps,preprint,superscriptaddress,showpacs]{revtex4-1}

\usepackage{graphicx}
\usepackage{dcolumn}
\usepackage{bm}

\begin{document}
\title{Photoinduced Kondo effect in CeZn$_{3}$P$_{3}$}

\author{J. Kitagawa}
\email{j-kitagawa@fit.ac.jp}
\affiliation{Department of Electrical Engineering, Faculty of Engineering, Fukuoka Institute of Technology, 3-30-1 Wajiro-higashi, Higashi-ku, Fukuoka 811-0295, Japan}
\author{D. Kitajima}
\affiliation{Department of Electrical Engineering, Faculty of Engineering, Fukuoka Institute of Technology, 3-30-1 Wajiro-higashi, Higashi-ku, Fukuoka 811-0295, Japan}
\author{K. Shimokawa}
\affiliation{Department of Electrical Engineering, Faculty of Engineering, Fukuoka Institute of Technology, 3-30-1 Wajiro-higashi, Higashi-ku, Fukuoka 811-0295, Japan}
\author{H. Takaki}
\affiliation{Department of Electrical Engineering, Faculty of Engineering, Fukuoka Institute of Technology, 3-30-1 Wajiro-higashi, Higashi-ku, Fukuoka 811-0295, Japan}

\date{\today}

\begin{abstract}
The Kondo effect, which originates from the screening of a localized magnetic moment by a spin-spin interaction, is widely observed in non-artificial magnetic materials, artificial quantum dots, and carbon nanotubes. In devices based on quantum dots or carbon nanotubes that target quantum information applications, the Kondo effect can be tuned by a gate voltage, a magnetic field, or light. However, the manipulation of the Kondo effect in non-artificial materials has not been thoroughly studied; in particular, the artificial creation of the Kondo effect remains unexplored. Per this subject study, however, a new route for the optical creation of the Kondo effect in the non-artificial material $p$-type semiconductor CeZn$_{3}$P$_{3}$ is presented. The Kondo effect emerges under visible-light illumination of the material by a continuous-wave laser diode and is ultimately revealed by photoinduced electrical resistivity, which clearly exhibits a logarithmic temperature dependency. By contrast, a La-based compound (LaZn$_{3}$P$_{3}$) displays only normal metallic behavior under similar illumination. The photoinduced Kondo effect, which occurs at higher temperatures when compared with the Kondo effect in artificial systems, provides a potential new range of operation for not only quantum information/computation devices but also for operation of magneto-optic devices thereby expanding the range of device applications based on the Kondo effect. 
\end{abstract}



\maketitle

\clearpage

\section{Introduction}
The Kondo effect has been widely studied in both bulk and thin-film magnetic materials\cite{Hewson:book1993,Andres:PRL1975,Sumiyama:JPSJ1986,Atac:Nanotech2013,Cha:Nanolett2010,Li:PRB2013,Lee:PRL2011,Lin:AMI2014}.
Contemporary developments in nanotechnology encourages the study of manipulating the Kondo effect through quantum dots (QDs) platforms \cite{Gordon:Nature1998,Cronenwett:Science1998} and carbon nanotubes (CNs)\cite{Nygard:Nature2000,Jarillo:Nature2005}; thus, this spurs additional applicability of the Kondo effect within the field of applied physics, particularly within the realm of quantum information and computation\cite{Nielsen:book2010}.
Recent pioneering works reporting the optical control of the Kondo interaction in QDs based on a nonlinear optical process\cite{Shahbazyan:PRL2000} and on manipulation of a localized spin\cite{Latta:Nature2011} inspired this study's team to investigate the possibility of the optical creation of the Kondo effect, with a focus on returning to non-artificial magnetic materials containing 4$f$-electrons. 
The Kondo energy scale ($\sim$ 10 to 100 K) in 4$f$-electron systems\cite{Hewson:book1993} is a few orders of magnitude higher than that ($\sim$ 10 mK to 1 K) of QDs\cite{Gordon:Nature1998,Cronenwett:Science1998} and CNs\cite{Nygard:Nature2000,Jarillo:Nature2005}, which is favorable for practical use of devices at higher temperatures.

\begin{figure*}
\begin{center}
\includegraphics[width=16cm]{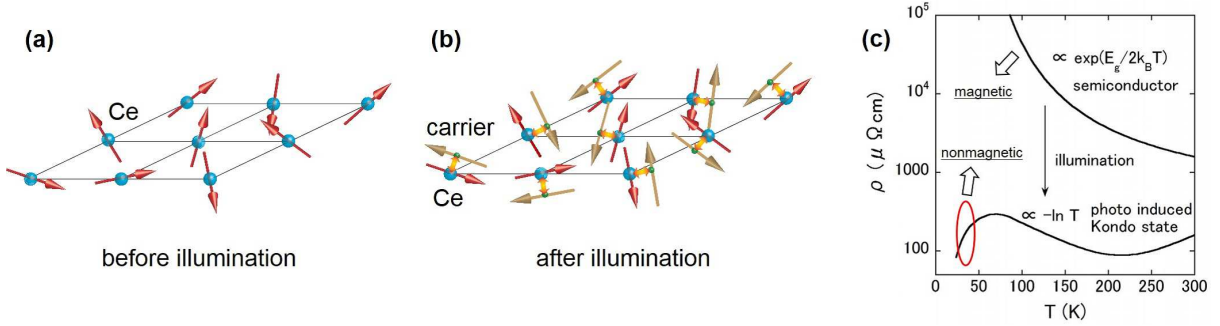}
\end{center}
\caption{(a) Ce-based semiconductor before illumination. Blue spheres and red arrows represent Ce atoms and 4$f$ electron magnetic moments, respectively. (b) Kondo coupling after illumination. Carriers are represented by green spheres. Brown arrows indicate carrier spin. (c) Expected temperature dependence of $\rho$(T) before and after illumination. }
\label{f1}
\end{figure*}

Figure 1(a) and 1(b) depict the proposed photoinduced Kondo effect. 
Figure 1(a) shows a semiconductor containing a Ce lattice with a well-localized 4$f$ magnetic moment at each lattice site. 
We eliminate conduction carriers, which are several orders of magnitude lower in number than the Ce atoms. 
Even if these conduction carriers and a proportion of the 4$f$ electrons can form a Kondo cloud, depopulation of the carrier concentration with decreasing temperature masks the emergence of the Kondo effect in observed transport properties; the temperature (T) dependence of the electrical resistivity, $\rho$(T), would exhibit semiconducting behavior represented by exp ($E_{g}/2k_{B}T$), where $E_{g}$ is the energy gap and $k_{B}$ is the Boltzmann constant. 
If above-band-gap excitation, for example, by a continuous-wave laser diode (CW LD), produces a sufficient number of quasi-equilibrium conduction carriers that exhibit strong Kondo interactions with Ce magnetic moments (as shown in Fig.\ 1(b)), then a Kondo screening can ultimately be created. 
Optical illumination can therefore change the property of a Ce-based magnetic material from a semiconducting-$\rho$ into a metallic-$\rho$ displaying the -lnT characteristic of the Kondo signature with, at low temperatures, Kondo coherence reflecting a Ce lattice structure (Fig.\ 1(c)).
Moreover, the unscreened 4$f$ spins govern the magnetic state of the semiconductor. 
By contrast, photoinduced Kondo screening leads to a nonmagnetic state, ascribed to via formation of spin singlets between localized spin and associated carriers.
The proposed mechanism for optical tuning of the Kondo effect is different from that reported in QDs\cite{Shahbazyan:PRL2000,Latta:Nature2011} and plays a major role in the study of wide-range carrier density tuning in the Kondo effect, which has not been comprehensively explored in long-term studies\cite{Wada:JPSJ1996,Haen:JLTP1987,Wohlleben:AP1985,Jaccard:PLA1992}.
From a practical viewpoint, optically controlled Kondo interactions may potentially lead to novel magneto-optical devices via modification of macroscopic magnetization, which would not be easily realized in QD (CN) devices because such devices require external metallic leads for an individual QD (CN). 

\begin{figure}
\begin{center}
\includegraphics[width=8cm]{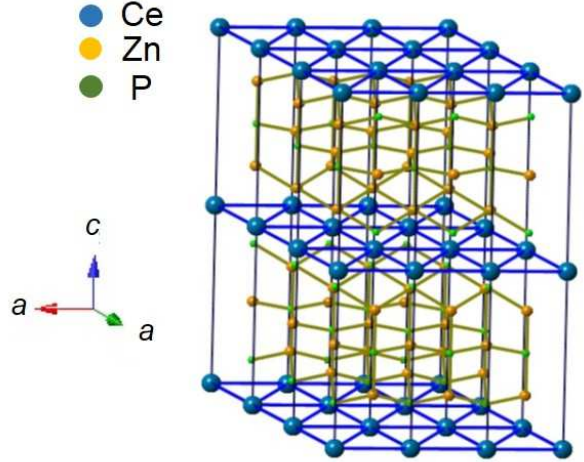}
\end{center}
\caption{3D view of CeZn$_{3}$P$_{3}$ ($a$=4.051 \AA, $c$=20.019 \AA). Blue, orange, and green spheres represent Ce, Zn, and P atoms, respectively.}
\label{f2}
\end{figure}

To implement the photoinduced Kondo effect in a simple experiment using a CW LD, a long photoexcited carrier lifetime ($\tau$) is required for the accumulation of sufficient carriers.
Recently, the study team observed photocarrier injection\cite{Kitagawa:JPSJ2013} into the semiconductor CeZn$_{3}$P$_{3}$ for which the value of $\tau$ was relatively high.
In our previous study\cite{Kitagawa:JPSJ2013} using a bulk sample, $\rho$ under visible-light illumination contained both $\rho$ of the original semiconductor and that of the photoexcited region because the optical penetration depth ($d_{p}$) was much smaller than the sample thickness. 
Because thin-film synthesis of CeZn$_{3}$P$_{3}$ is generally quite challenging to execute, a $\rho$(T) extraction method for the photoexcited region was hence developed in a rigorous manner by checking $\rho$(T) reproducibility under illumination and by using supplemental data for analysis. 
CeZn$_{3}$P$_{3}$ crystallizes into hexagonal ScAl$_{3}$C$_{3}$-type\cite{Nientiedt:JSSC1999}, as shown in Fig.\ 2. 
Moreover, Ce atoms form a triangular lattice of a two-dimensional layer, via the sandwiching of Zn and P atoms. 
Curie-Weiss Law behavior in the magnetic susceptibility realm indicated localized 4$f$ spins of the Ce$^{3+}$ ions\cite{Yamada:JPCS2010}. 
Reflecting such a frustrated Ce atom geometry, CeZn$_{3}$P$_{3}$ resultantly displayed a low magnetic ordering temperature\cite{Yamada:JPCS2010} of 0.75 K. 
Therefore, in accordance with the Doniach phase diagram\cite{Hewson:book1993,Brandt:AP1984} for a compound with weakened magnetic interaction between 4$f$ spins, the Kondo interaction dominates the photoexcited system.

\section{Materials and Methods}
Polycrystalline CeZn$_{3}$P$_{3}$, LaZn$_{3}$P$_{3}$, and LaZn$_{3}$P$_{2.99}$Si$_{0.01}$ samples were synthesized by a solid-state reaction technique using Ce (or La) pieces (99.9\%), Zn shot (99.99\%), red P powder (99.999\%), and Si powder (99.999\%).
The study team employed the use of CeZn$_{3}$ or LaZn$_{3}$, prepared by direct reaction of the constituent elements in an evacuated quartz tube that was heated at 850 $^{\circ}$C for 1 h as a solid-state reaction precursor. 
Crushed CeZn$_{3}$ (LaZn$_{3}$) and P (and Si) with a molar ratio CeZn$_{3}$ (LaZn$_{3}$):P = 1:3 or LaZn$_{3}$:P:Si=1:2.99:0.01 were homogeneously mixed together in a glovebox. 
The pelletized sample was then reacted in an evacuated quartz tube at 850 $^{\circ}$C for 2 days, and subsequently evaluated using a powder X-ray diffractometer with Cu-K$\alpha$ radiation and virtually a single phase. 

The team then checked the optical band gaps of CeZn$_{3}$P$_{3}$ and LaZn$_{3}$P$_{3}$ via their diffuse reflectance $R_{d}$ spectra and recorded them using a UV-Vis spectrometer (Shimadzu, UV-3600). 
Moreover, associated band calculations were carried out using the WIEN2K package\cite{Blaha:book2001}, and the extinction constant was measured using a variable-angle spectroscopic ellipsometer (J. A. Woollam, M-2000).
$\rho$(T) values between 30 and 300 K under illumination were measured by the conventional DC four-probe method using a closed-cycle He gas cryostat.
A parallelepiped sample with dimensions of $\sim$1$\times$1.5$\times$10 mm$^{3}$ was cut from a pellet of each sample, and the distance between the voltage electrodes was 2$\sim$3 mm.
All electrodes were covered by thin metal plates to reduce any potential extrinsic photovoltaic effects. 
The optical source was an LD with a photon energy of 1.85 eV, which was sufficient for excitation of electrons across the band gap.
The light beam from the LD was focused onto the area between the voltage electrodes. 

The relaxation time for the photoexcited carriers was estimated by measuring the photovoltage time dependence of the sample used in the electrical resistivity measurements with photovoltages ultimately recorded via the use of a nanovoltmeter (Keithley, 181). 
The settling time for the voltage range used was 0.5 s, which was shorter than the photocarrier lifetime of the sample under test. 
Hall effect measurements were performed by the Van der Pauw method under a magnetic field of 0.5 T produced by a variable DC electromagnet (Tamagawa, TM-YSF8615C-083). 

\section{Results and discussion}

\begin{figure}
\begin{center}
\includegraphics[width=10cm]{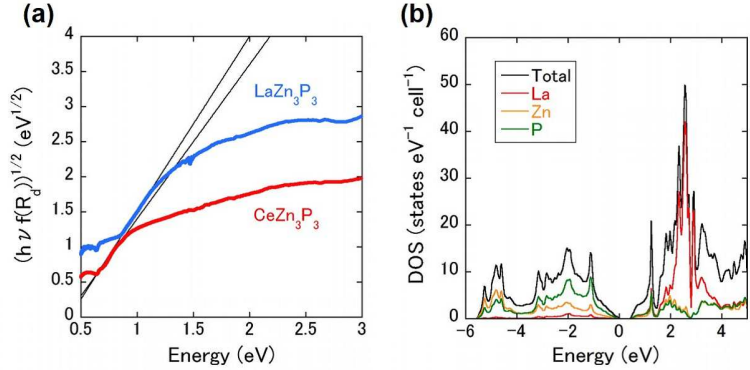}
\end{center}
\caption{(a) Spectra of $(h\nu f(R_{d}))^{1/2}$ of CeZn$_{3}$P$_{3}$ and La-based counterpart. (b) Density of states per unit cell of LaZn$_{3}$P$_{3}$. The black curve represents the total density of states. Red, orange, and green curves represent the projected density of states of the individual La, Zn, and P atoms, respectively.}
\label{f3}
\end{figure}

Figure 3 (a) shows the spectra of $(h\nu f(R_{d}))^{1/2}$ of CeZn$_{3}$P$_{3}$ and the La-based counterpart. 
$R_{d}$ is the diffuse reflectance.
The Kubelka-Munk function\cite{Marx:PRB1984} $f(R_{d})$, corresponding to the absorption coefficient, can be calculated using the equation of $f(R_{d})=\frac{(1-R_{d})^2}{2R_{d}}$.
$f(R_{d})$ is related to $E_{g}$ through $(h\nu f(R_{d}))^{1/m}=A(h\nu-E_{g})$, where $h$, $\nu$, and $A$ are the Planck constant, the frequency of the light, and a proportionality factor, respectively\cite{Tauc:PSS1966}.
The value of $m$ depends on the optical transition type. 
The study team used $m$=2, following the report on the indirect-gap isostructural-semiconductor\cite{Tejedor:JCG1995} SmZn$_{3}$P$_{3}$ and also the ab-initio band calculation results for LaZn$_{3}$P$_{3}$.
As shown in Fig.\ 3(a), $E_{g}$ of 0.37 (0.40) eV was determined using the tangential line for CeZn$_{3}$P$_{3}$ (LaZn$_{3}$P$_{3}$). 
For each compound, the spectral weight at lower energies deviating upwards from the solid line indicates the existence of impurity states in the energy gap.

Fig.\ 3(b) shows the total density of states (DOS) profile of LaZn$_{3}$P$_{3}$ with the Fermi energy ($E_{F}$) located at 0 eV. 
The projected DOS profiles for the La, Zn, and P atoms are also provided. 
$E_{g}$ was estimated to be 0.4 eV, which agrees well with the experimentally determined value from the $R_{d}$ spectrum. 
The valence and conduction bands are primarily composed of orbitals with covalent bond characteristics between the Zn and P atoms and the La 5$d$ orbital. 
Experimental trials derived $d_{p}$ at 1.85 eV (which is the photon energy of the LD) from the extinction constant as 44 nm for CeZn$_{3}$P$_{3}$ (and 47 nm for LaZn$_{3}$P$_{3}$).
At 1.85 eV and probably in both compounds, an optical transition from the covalent Zn-P orbitals to the empty La (Ce) 5$d$ state and/or the antibonding Zn-P orbitals would be induced. 

\begin{figure}
\begin{center}
\includegraphics[width=16cm]{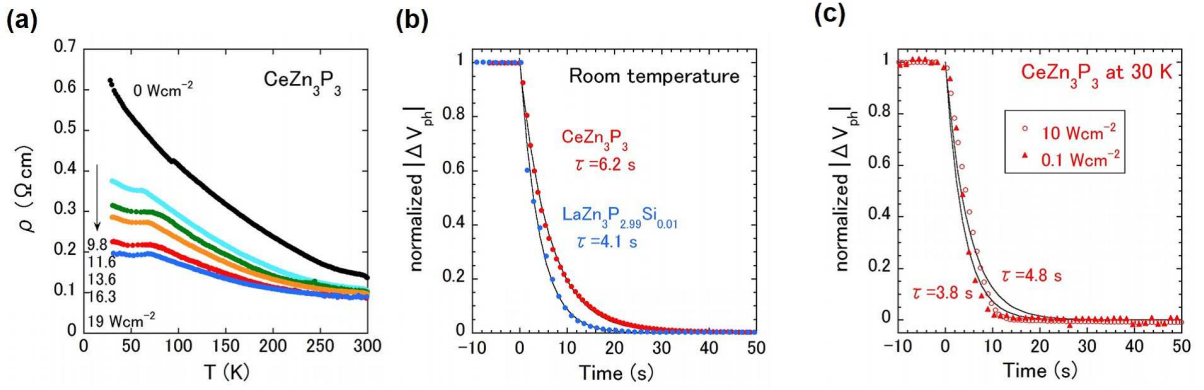}
\end{center}
\caption{(a) Temperature dependence of $\rho$(T) of CeZn$_{3}$P$_{3}$ at 0, 9.8, 11.6, 13.6, 16.3, and 19 Wcm$^{-2}$. (b) Time-evolution of decaying normalized $\left|\Delta V_{ph} \right|$ after illumination ($\sim$ 10 Wcm$^{-2}$) ceases for CeZn$_{3}$P$_{3}$ (red mark) and LaZn$_{3}$P$_{2.99}$Si$_{0.01}$ (blue mark) at room temperature. (c) Time-evolution of decaying normalized $\left|\Delta V_{ph} \right|$ after illumination ($\sim$ 10 Wcm$^{-2}$ or 0.1 Wcm$^{-2}$) ceases at 0 s for CeZn$_{3}$P$_{3}$. The temperature is 30 K.}
\label{f4}
\end{figure}

Figure 4(a) shows $\rho$(T) for CeZn$_{3}$P$_{3}$ at several fluence rates. 
With increasing fluence rate, while the semiconducting behavior is retained, systematic depression of $\rho$ is observed, indicating carrier injection by visible light. 
It should be noted that the difference of $\rho$(T) between the dark and the illuminated state is sufficiently large.
This means that the photoinduced electrical resistivity is not strongly affected by $\rho$(T) of the dark state\cite{SM1}.
Figure 4(b) shows the normalized absolute value of photovoltage change $\left|\Delta V_{ph} \right|$.
The time dependence of $\left|\Delta V_{ph} \right|$ for each compound at room temperature measured after the light ($\sim$ 10 Wcm$^{-2}$) was switched off at 0 s, depicts a single exponential decay scenario as is indicated by the solid curve. 
Slight hole doping was conducted for LaZn$_{3}$P$_{3}$. 
It was found that virtually singular exponentially decaying curves persist for both compounds down to low temperatures. 
The photoinduced carrier concentration $n$ reaches the order of 10$^{23}$ cm$^{-3}$ using $n=F\tau/h\nu d_{p}$, where $F$ is the fluence rate in Fig.\ 4(a), and metallization is anticipated due to illumination via the CW LD.
However, because $d_{p}$ at 1.85 eV (44 nm) is much smaller than the sample thickness (1 mm), metallization would not occur throughout the sample volume.

Thermal effects on $\tau$ were ultimately verified, especially at lower temperatures. 
Figure 4(c) shows the comparison of time traces of normalized $\left|\Delta V_{ph} \right|$ at 30 K, for CeZn$_{3}$P$_{3}$, at $F$ between 10 Wcm$^{-2}$ and 0.1 Wcm$^{-2}$. 
It was determined that the latter $F$ is too weak to expect noticeable thermal effects on $\tau$. 
Although a slight increase in $\tau$ occurs by increasing the value of $F$, which indicates that a weak thermal effect certainly exists, a long $\tau$ on the order of a second is ultimately obtained (even at 0.1 Wcm$^{-2}$), and is essentially intrinsic for the compound realizing the metallization process. 
The fluence rate dependence of $\tau$ indicates that the overall thermal effect is weak.
As such, the study team moreover estimated the temperature increase, $\Delta$T, due to the thermal effect. 
The sample was placed on a large cooling stage under continuous CW LD illumination. 
Next, $\Delta$T was calculated as follows under equilibrium conditions: if all of the light power with fluence rate $F$ was transformed into heat within a sample, $\Delta$T could be expressed by $\frac{Fd_{s}}{\kappa}$, where $\kappa$ and $d_{s}$ are the thermal conductivity and thickness of the sample, respectively.
If we use the value of 50 W/mK for $\kappa$, which is the value of a typical semiconductor\cite{Geballe:PR1958} at 30 K, then $\Delta$T at that temperature under 20 Wcm$^{-2}$ $F$ reaches 4 K, which is comparable to the value cited in related literature value\cite{Petrakovskii:JETP1986}.

An inhomogeneous carrier distribution in a quasi-equilibrium state is produced in the direction of the sample depth.
Although the ideal carrier distribution would show an exponential decay, diffusion or surface recombination processes result in a carrier distribution that is of a non-exponential decay type.
The carrier concentration ($n (z)$) in the sample depth direction for the quasi-equilibrium state is obtained using a one-dimensional diffusion equation as follows\cite{Vaitkus:PSSA1976}:
\begin{equation}
\frac{\partial n(z,t)}{\partial t}=D\frac{\partial^{2} n(z,t)}{\partial z^{2}}-\frac{n(z,t)}{\tau}+g\delta(t)\exp\left(-\frac{z}{d_{p}}\right),
\label{equ:1-diffusion}
\end{equation}
where $n(z, t)$ depends on the time ($t$) and position ($z$) in the sample depth, and $g$ and $\delta$(t) are the factor corresponding to the fluence rate and the $\delta$-function, respectively. 
$D$ is the diffusion coefficient and is given by
\begin{equation}
D=\frac{\mu_{bi}k_{B}T_{s}}{|e|},
\label{equ:diffcoeff}
\end{equation}
where 1/$\mu_{bi}$ is equal to 1/$\mu_{e}$+1/$\mu_{h}$ and $T_{s}$ is the sample temperature. 
The solution\cite{Vaitkus:PSSA1976} of the diffusion equation is thus given by:
\begin{eqnarray}
n(z,t)&=&g\exp\left(-\frac{z^{2}}{4Dt}\right)\left\{\frac{1}{2}\left[h\left(\frac{\sqrt{Dt}}{d_{p}}-\frac{z}{2\sqrt{Dt}}\right)+\frac{\frac{D}{d_{p}}+v_{s}}{\frac{D}{d_{p}}-v_{s}}h\left(\frac{\sqrt{Dt}}{d_{p}}+\frac{z}{2\sqrt{Dt}}\right)\right]\right. \nonumber \\
&&\left.-\frac{v_{s}}{\frac{D}{d_{p}}-v_{s}}h\left(v_{s}\sqrt{\frac{t}{D}}+\frac{z}{2\sqrt{Dt}}\right)\right\}\exp\left(-\frac{t}{\tau}\right),
\label{equ:diff-sol}
\end{eqnarray}
where $v_{s}$ is the surface recombination velocity, and $h(z)$ is related to the error function by $h(z)=\exp(z^{2})(1-$erf$(z))$. 
The carrier number in the quasi-equilibrium state requires the integration of $n(z, t)$ with respect to $t$; i.e.,
\begin{equation}
n(z)=\int_0^\infty n(z,t)dt.
\label{equ:carnumber}
\end{equation}

\begin{figure}
\begin{center}
\includegraphics[width=6cm]{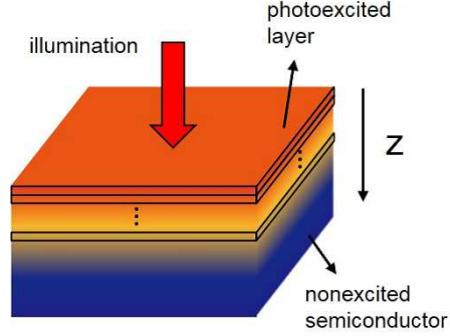}
\end{center}
\caption{Schematic of photoexcited semiconductor in the equilibrium state. The photoexcited layer is depicted in light orange. The inhomogeneity of the concentration of injected and/or diffused quasi-equilibrium carriers in the sample depth $z$ are expressed by color gradation. The blue color represents the non-excited region. The thin platelet with the black frame is a discretized layer with a thickness much smaller than that of the optical penetration depth. The carrier concentration is assumed to be homogeneous in each platelet. Here, we note that the thickness of the photoexcited layer ($\sim$ 50 nm) is much smaller than the sample thickness ($\sim$ 1 mm).}
\label{f5}
\end{figure}

As shown in Fig.\ 5, $n(z)$ is discretized in layers that are much thinner than $d_{p}$.
By following the parallel circuit model\cite{Petritz:PR1956,Okazaki:JPSJ2012}, which is applied to the discretized photoexcited structure, the electrical conductivity of the entire sample volume $\sigma_{p}$ can be described as follows:
\begin{equation}
\sigma_{p}=\frac{t_{p}}{d_{s}}\sum_{k=1}^l\sigma_{k}+\left(1-\frac{lt_{p}}{d_{s}}\right)\sigma_{0},
\label{equ:sigma_p}
\end{equation}
where $t_{p}$ is the layer thickness of a discretized platelet, and $\sigma_{k}$ and $\sigma_{0}$ are the electrical conductivities of each platelet and that of the non-excited region, respectively. 
By introducing the maximum carrier concentration ($n_{max}$) and the corresponding maximum conductivity ($\sigma_{max}$), the excess conductivity due to photo-illumination ($\sigma_{k}-\sigma_{0}$) can be approximated by
\begin{equation}
\sigma_{k}-\sigma_{0}=\sigma_{max}\frac{n_{k}}{n_{max}}.
\label{equ:excess_sigma}
\end{equation}
Next, the photoinduced electrical resistivity ($\rho_{photo}$) at the highest carrier concentration is calculated as $\sigma_{max}^{-1}$, whereby
\begin{equation}
\sigma_{max}=\frac{d_{s}}{t_{p}}\frac{n_{max}}{\sum_{k=1}^ln_{k}}(\sigma_{p}-\sigma_{0}).
\label{equ:sigma_max}
\end{equation}

The diffusion equation, which determines $n(z)$, has quantitatively explained photoinduced optical conductivities in conventional bulk semiconductors, such as GaAs\cite{Beard:PRB2000} and Si\cite{Kitagawa:JPCM2007}. 
The parallel-circuit model, with no consideration of an inhomogeneous $n(z)$, can also extract a rather accurate photoinduced DC conductivity value in a thick sample\cite{Okazaki:JPSJ2012}. 
Note that the developed model is a combination of two models, both of which are applicable for the case of a $d_{p}$ that is much smaller than a given sample thickness. 
Hall effect measurements showed that the material was $p$-CeZn$_{3}$P$_{3}$, with a $\mu_{h}$=7.2 cm$^{2}$V$^{-1}$s$^{-1}$. 
Although $\mu_{e}$ should be included when determining $\mu_{bi}$, only $\mu_{h}$ was used to extract $\rho_{photo}$(T) because the hole carriers govern the Kondo interaction with the 4$f$ electrons\cite{Mahan:book1998}. 
However, calculation of $\mu_{bi}$ with an equal $\mu_{e}$ contribution does not greatly alter $\rho_{photo}$(T)\cite{SM2}.

\begin{figure*}
\begin{center}
\includegraphics[width=16cm]{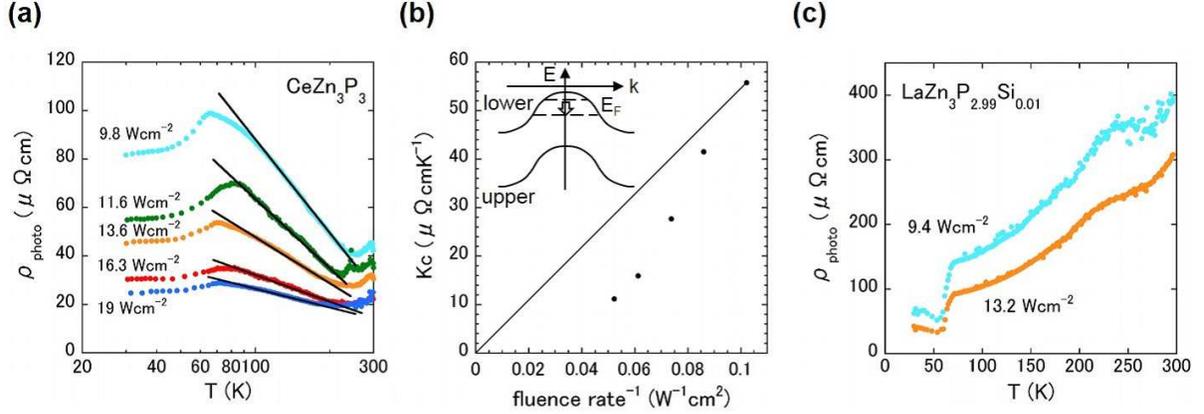}
\end{center}
\caption{(a) Temperature dependence of $\rho_{photo}$(T) of CeZn$_{3}$P$_{3}$ extracted at 9.8, 11.6, 13.6, 16.3, and 19 Wcm$^{-2}$. (b) Inverse fluence rate dependence of $K_{c}$, which is a proportionality factor for -lnT. The solid line shows the predicted relationship between $K_{c}$ and the inverse fluence rate normalized at 9.8 Wcm$^{-2}$ with the assumption that only the carrier concentration relies on the fluence rate. The inset shows a schematic of the hybridization band. (c) Temperature dependence of $\rho_{photo}$(T) of LaZn$_{3}$P$_{2.99}$Si$_{0.01}$ extracted at 9.4 and 13.2 Wcm$^{-2}$. }
\label{f6}
\end{figure*}

Figure 6(a) shows the extracted $\rho_{photo}$(T), with values indicating remarkably metallic behavior. 
As the temperature is reduced from 300 K to approximately 200 K, each $\rho_{photo}$ characteristic displays a positive slope, which also suggests a metallic state. 
However, with further temperature reduction, all data show an increase in $\rho_{photo}$ according to the -lnT Kondo signature shape (see the solid lines in Fig.\ 6(a)). 
At approximately 70 K, each $\rho_{photo}$ characteristic shows a broad hump, which may be caused by a crystalline-electric-field effect or a photoinduced phase transition. 
It was noted at this instance that the La-based counterpart also shows an anomaly at similar temperatures; this phenomenon is further explained below. 
Some carrier-localization effects might potentially be attributable to the upturn of $\rho_{photo}$ below 200 K in CeZn$_{3}$P$_{3}$, which is discussed in Appendix A.
Hall effect measurements indicated hole conduction in CeZn$_{3}$P$_{3}$.
The Kondo effect in Ce compounds is generally induced by Kondo interaction between 4$f$ electrons and hole carriers\cite{Hewson:book1993}. 
The study team therefore proposes that photo-illumination produces more hole carriers that interact with Ce 4$f$ electrons, with separated electrons ultimately providing a negligible contribution to -lnT dependence.
It was determined that the -lnT dependence of the metallic state supported by a low $\rho$-value demonstrates the occurrence of the photoinduced Kondo effect.
It should be remarked that the large photoconductivity is not intimate with the conventional theory, but the comment on this problem is given in Appendix B.
Moreover, in the case of polycrystalline samples, the photoconductivity due to grain boundary effects is important, and is discussed in detail in Appendix C.

It is interesting to investigate a transition from $\rho_{photo}$(T) without the Kondo effect to that of the Kondo metal, by measuring $\rho_{photo}$(T) at a low fluence rate.
Although the preliminary $\rho_{photo}$(T) at 1.7 Wcm$^{-2}$ supports the possible thermal activation type, the reliability is rather low due to the small difference of $\rho$(T) between the dark and the illuminated state.
The considerable reduction of $\rho$(T) under the illumination is necessary to obtain the reliable $\rho_{photo}$(T), which is inevitably of metallic state. 
To investigate the crossover from the semiconductor to the Kondo metal, we need a thin film sample, which does not require the model analysis to extract $\rho_{photo}$(T). 

Figure 6(a) offers a novel opportunity to consider $n$ dependence of the Kondo interaction. 
Long-term studies on heavy fermions have revealed the magnetic-impurity concentration dependence\cite{Sumiyama:JPSJ1986} of $\rho$(T) or pressure dependence\cite{Wohlleben:AP1985} of $\rho$(T) in which the hybridization between the carrier and the 4$f$ electron is tuned.
In both cases, $n$ is kept constant.
The $n$ dependence of -lnT under a constant magnetic-ion concentration is non-trivial.
The proportionality factor $K_{c}$ of -lnT, gradually decreases with increasing fluence rate which is proportional to $n$. 
As such, inverse fluence-rate dependencies of $K_{c}$ were plotted in Fig.\ 6(b). 
By assuming that the simple relation $\rho$ = $m^{*}/ne^{2}\tau_{k}$ applies, where $m^{*}$ is the effective mass of the carriers and $\tau_{k}$ is the momentum relaxation time, one can draw a solid line representing the proportional relationship between $1/n$ and $K_{c}$, normalized at 9.8 Wcm$^{-2}$. 
The downward deviation from the solid line shows that $\rho_{photo}$ is indeed also affected by $m^{*}$, in addition to $n$.
At room temperature, $n$ of CeZn$_{3}$P$_{3}$ determined by Hall effect measurements is 2.3$\times$10$^{18}$ cm$^{-3}$, from which $n$ for the holes at 9.8 Wcm$^{-2}$ (19 Wcm$^{-2}$) is estimated to be 2.8$\times$10$^{21}$ cm$^{-3}$ (5.6$\times$10$^{21}$ cm$^{-3}$) by considering $n$ as proportional to the conductivity, wherein both excited holes and electrons equally participate. 
The calculated Ce atom concentration is 7$\times$10$^{21}$ cm$^{-3}$. 
Therefore, if $n$ originates from a Kondo lattice hybridization band (see the inset of Fig.\ 6(b)), then the value of $n_{h}+n_{f}$ (i.e., the full occupancy of the lower band: 2.0), where $n_{h}$ and $n_{f}$ are the numbers of holes and 4$f$ electrons per Ce site, respectively, is 0.4 (0.8) at 9.8 Wcm$^{-2}$ (19 Wcm$^{-2}$) and $E_{F}$ at 19 Wcm$^{-2}$ may then be ascertained as located near the top of the lower hybridization band. 
As $E_{F}$ is reduced from the top of the band by optical pumping, continuous $m^{*}$ reduction would be expected until a dispersion-free $f$-character band appears, which qualitatively explains the downward deviation portrayed in Fig.\ 6(b). 
Moreover, one must also consider the $n$ dependence of $\tau_{k}$, which requires a realistic theoretical treatment of a strongly-correlated electron system.
In this study, the estimated range of $n$ (2.8$\times$10$^{21}$ $\sim$ 5.6$\times$10$^{21}$ cm$^{-3}$) may be narrow, therefore leading to a small change of the Kondo temperature under the illumination.
This would be the reason of no clear fluence rate dependence of photoinduced Kondo effect except the fluence rate dependence of $K_{c}$.

Further strong evidence of the photoinduced Kondo effect in CeZn$_{3}$P$_{3}$ is the lack of occurrence of the Kondo effect in photoexcited LaZn$_{3}$P$_{3}$. 
Because the non-doped $p$-LaZn$_{3}$P$_{3}$ shows a high $\rho$ value, light Si doping is performed to assist hole conduction, as in LaZn$_{3}$P$_{2.99}$Si$_{0.01}$. 
Using $\mu_{bi}$ = 6.8 cm$^{2}$V$^{-1}$s$^{-1}$ as determined by Hall effect measurements and $\tau$ = 4 s obtained per Fig.\ 4(b), $\rho_{photo}$ at the highest carrier concentration is thereby extracted, as shown in Fig.\ 6(c), indicating sufficient metallization with normal metal temperature dependence. 
The sharp drops in $\rho_{photo}$ below 70 K are possibly due to a phase transition enhanced by the illumination (not shown). 
The shorter $\tau$ in LaZn$_{3}$P$_{2.99}$Si$_{0.01}$ is responsible for the rather high $\rho_{photo}$ when compared to that of CeZn$_{3}$P$_{3}$. 

Thus, the photoinduced Kondo effect is sustained by realization of the metallic state due to illumination by the CW LD. 
The metallization stems from the long separation times of the electron-hole pairs. 
The results of photocatalyst research\cite{Chen:CR2010} indicate that the associated anisotropic crystal structure of this scenario forms a rather large internal dipole field, contributing to the long electron-hole pair separation time.
The crystal structure of CeZn$_{3}$P$_{3}$ is actually hexagonal, with a highly anisotropic $c/a$ ratio (see Fig.\ 2). 
Note that the effect of grain boundaries frequently contributes to the persistent levels of photoconductivity\cite{Shalish:PRB2000}. 
Hence, the grain boundary may be another source of origin prompting the ostensibly long $\tau$. 

\section{Conclusions}

This subject study has observed and resultantly interpreted the photoinduced Kondo effect in $p$-type CeZn$_{3}$P$_{3}$. 
Metallization due to illumination via the CW LD is caused by a long $\tau$ and is attributable to the highly anisotropic CeZn$_{3}$P$_{3}$ crystal structure and/or grain boundary. The Kondo effect emerges under visible-light illumination of the subject material by a CW LD and is ultimately revealed by photoinduced electrical resistivity conditions to clearly exhibit a logarithmic temperature dependency. By contrast, a La-based compound (LaZn$_{3}$P$_{3}$) under similar illumination displays only normal metallic behavior. 
 
These subject findings raise new fundamental issues with regard to the crossover from a non-correlated electron system to a strongly correlated system along with its associated quantum entanglement formation process. Experimental results obtained using a CW LD may be advantageous for low-cost practical applications in next-generation magneto-optic and quantum information/computation devices. In essence, the photoinduced Kondo effect, which occurs at higher temperatures compared with the Kondo effect in artificial systems, can notably provide a potential new range of functionalities within a litany of devices that operate based upon the Kondo effect principle.

\begin{acknowledgments}
The study team is grateful for the financial support provided by the Asahi Glass Foundation and the Fukuoka Institute of Technology's Comprehensive Research Organization.
\end{acknowledgments}

\section{APPENDIX A}
If the localization effect is operative, then both Ce and La compounds would be expected to show an upturn of $\rho_{photo}$. 
In carrier localization, the degree of disorder affects the position of $T_{min}$, below which $\rho$ starts to increase\cite{Bhosle:JAP2006}.
The tuning of carrier density by various illumination-power levels changes the degree of disorder, resulting in a shift of $T_{min}$;
in fact, such a shift of $T_{min}$ under varying degrees of illumination has been previously reported in a semiconductor\cite{Kozasa:MRE2014}.
Hence, the lack of optical dependence of $T_{min}$ in this study's results strongly indicates that carrier localization would likely not be responsible for the upturn of $\rho_{photo}$ in CeZn$_{3}$P$_{3}$. 

\section{APPENDIX B}
As mentioned in the main text, the estimated value of $n$ for the holes at 9.8 Wcm$^{-2}$ (19 Wcm$^{-2}$) is 2.8$\times$10$^{21}$ cm$^{-3}$ (5.6$\times$10$^{21}$ cm$^{-3}$).
The conventional theory of photoconductivity predicts that conductivity saturates with increasing pump power due to the full occupation of trapping sites in the band gap\cite{Bube:JAP1966}. 
Because the density of traps is generally limited in the range of approximately 10$^{16}$ cm$^{-3}$ to 10$^{19}$ cm$^{-3}$, $n$ of 10$^{21}$ cm$^{-3}$ per this study, such a phenomenon cannot be explained by the conventional photoconductivity theory. 
However, the excitation photon energy (1.85 eV) is far above $E_{g}$ (0.37 eV), which is not directly taken into account by the theory. 
CeZn$_{3}$P$_{3}$ with a rather low $\rho$ ($\sim$0.1 $\Omega$cm) would approximate the conditions of a degenerated $p$-semiconductor, suggesting an $E_{F}$ near the top of the valence band. 
Moreover, $E_{F}$ can easily move below the top of the valence band with increasing optical power, as mentioned in the main text, which is also not substantiated by the subject photoconductivity theory assuming an $E_{F}$ within the band gap. 
Thus, the overall applicability of the conventional photoconductivity theory is duly questionable within the scope of this study. 
Additional influential candidates within the realm of carrier trapping, however, were acknowledged during the assessment process. 
One associated candidate was the grain boundary\cite{Shalish:PRB2000}, while the other candidate was the energy-band grouping far above the lowest conduction band level. 
The density of states for the latter is resultantly sufficient to sustain a carrier concentration of 10$^{21}$ cm$^{-3}$ (see also Fig. 3(b)).

\section{APPENDIX C}
The potential effects of grain boundaries were likewise evaluated, which are additional likely catalysts for increases in photoconductivity\cite{Petritz:PR1956}.
If $\rho_{photo}$(T) due to the photoinduced lowering of a grain-boundary potential-barrier shows an upturn below a certain temperature, it was hence found to approximately follow a thermal activation-type trend\cite{Shalish:PRB2000}. 
Therefore, $\rho_{photo}$ produced by only grain boundary effects cannot explain $\rho_{photo}$(T) of CeZn$_{3}$P$_{3}$ showing an -lnT behavior. 
Moreover, the $\rho_{photo}$ of each compound is of the order of $\mu$$\Omega$cm, with the difference of $\rho_{photo}$(T) between CeZn$_{3}$P$_{3}$ and LaZn$_{3}$P$_{3}$ being irrelevant to the existence of grain boundaries due to both polycrystalline compounds possessing grain boundaries themselves. 
These facts suggest that the possibility of $\rho_{photo}$ for CeZn$_{3}$P$_{3}$ being dominated by the grain boundary effect is unlikely. 
Although the grain boundary effect would not dominate $\rho_{photo}$(T) of CeZn$_{3}$P$_{3}$, the effect may nevertheless play a certain role in the emergence of the Kondo effect. 
As such, at the present stage, the overall potential effects from grain boundaries may still hold a certain degree of discernibility. 
Single-crystalline thin-film fabrication of CeZn$_{3}$P$_{3}$ is highly desired in support of experimental trials geared toward the assessment of a photoinduced Kondo effect. 
If the subject CeZn$_{3}$P$_{3}$ thin-film does not ultimately show the photoinduced effect, it can then be definitively asserted that the grain boundary significantly contributes to the emergence of the Kondo effect.

\end{document}